\documentclass{elsart}

\begin{document}

\begin{frontmatter}

\title{Dynamic Asymmetric Communication}

\author{Travis Gagie}
\address{Department of Computer Science\\
        University of Toronto}

\begin{abstract}
We show how any dynamic instantaneous compression algorithm can be
converted to an asymmetric communication protocol, with which a server
with high bandwidth can help clients with low bandwidth send it messages.  
Unlike previous authors, we do not assume the server knows the messages'
distribution, and our protocols are the first to use only one round of
communication for each message.
\end{abstract}

\begin{keyword}
Data compression; asymmetric communication.
\end{keyword}

\end{frontmatter}

\section{Introduction}
\label{introduction}

Internet users usually download more than they upload, and many
technologies have asymmetric bandwidth --- greater from servers to clients
than from clients to servers.  Adler and Maggs~\cite{AM01} considered
whether a server can use its greater bandwidth to help clients send it
messages.  They proved it can, assuming it knows the messages'
distribution.  We argue this assumption is often both unwarranted and,
fortunately, unnecessary.

Suppose a number of clients want to send messages to a server.  At any
point, the server knows all the messages it has received so far; each
client only knows its own messages and does not overhear communication
between other clients and the server.  Thus, the server may be able to
construct a good code but the clients individually cannot.  Adler and
Maggs assumed the server, after receiving a sample of messages, can
accurately estimate the distribution of \emph{all} the messages.  This
assumption let them simplify the problem: Can the server help a
\emph{single} client send it a message drawn from a distribution known to
the server?  Given a representative sample of messages and a protocol for
this simpler problem, the server can just repeat the protocol for each
remaining message.  In fact, it can even do this in parallel.

Adler and Maggs gave protocols for the simpler problem in which the server
uses its knowledge to reduce the expected number of bits the client sends
to roughly the entropy of the distribution.  Their work has been improved
and extended by several authors~\cite{WAF01,GGS01,LH02,BKLM03}, whose
results are summarized in
Table~\ref{results},\footnote{Table~\ref{results} does not include a
recent paper by Adler~\cite{Adl05}, in which he considered a harder
version of the original problem with many clients: Can the server take
advantage of correlations between messages?  He showed it can, but used
the even stronger assumption that the server knows the probability
distribution over entire sequences of messages.} and used in the Infranet
anti-censorship system~\cite{FBHBK02,FBWBK03}.  However, while
implementing Infranet, Wang~\cite{Wan03} found the distribution of the
messages (webpage requests) changed over time --- the sample was
unreliable.

\begin{table}[ht]
\begin{center}
\begin{tabular}{c|c|c|c}
References \hspace{.5ex}
	& \hspace{.5ex} Bits sent by Server \hspace{.5ex}
	& Bits sent by Client
	& Rounds\\ \hline
\cite{AM01,LH02}
	& \(3 \lceil \log n \rceil\)
	& \(1.09 H + 1\)
	& \(1.09H + 1\)\\
\cite{AM01}
	& \(O (\log n)\)
	& \(O (H + 1)\)
	& \(O (1)\)\\
\cite{WAF01}
	& \((H + 2) \lceil \log n \rceil\)
	& \(H + 2\)
	& \(H + 2\)\\
\cite{WAF01}
	& \(O (2^k H \log n)\)
	& \(H + 2\)
	& \hspace{.5ex} \((H + 1) / k + 2\)\\
\cite{GGS01}
	& \(k H \lceil \log n \rceil + 1\)
	& \(H \log_{k - 1} k + 1\)
	& \(H / \log k + 1\)\\
\cite{BKLM03}\rule{0ex}{2.5ex}
	& \((k + 2) \lceil \log n \rceil\)
	& \hspace{.5ex}
	\(\frac{H \log (k + 2)} {\log (k + 2) - 1} + \log (k + 2)\)
	\hspace{.5ex}
	& \(\frac{H}{\log (k + 2) - 1} + 1\)
\end{tabular}
\end{center}
\caption{Suppose a server tries to help a client send it one of $n$ 
messages, chosen according to a distribution with entropy $H$ that is 
known to the server but not the client.  Adler and Maggs~\cite{AM01}, 
Watkinson, Adler and Fich~\cite{WAF01}, Ghazizadeh, Ghodsi and 
Saberi~\cite{GGS01} and Bose, Krizanc, Langerman and Morin~\cite{BKLM03} 
gave protocols for this problem with expected-case upper bounds as shown; 
the last three protocols take a parameter \(k \geq 1\).  This table is 
based on one given by Bose {\it et al.}\ but we use a different notation.}
\label{results}
\end{table}

\section{Dynamic compression and asymmetric communication}
\label{compression}

In this section we show how algorithms for dynamic instantaneous data
compression --- e.g., dynamic Huffman coding~\cite{Knu85} and
move-to-front compression~\cite{BSTW86} --- can be converted to dynamic
asymmetric communication protocols.  By dynamic protocols, we mean ones
not needing the server to know the messages' distribution.

For dynamic instantaneous compression (also called prefix-free
compression), an encoder makes a single pass over a string $S$ and writes
a codeword after reading each character; a decoder can later make a single
pass over the encoding and writes a character of $S$ after reading each
codeword.  Implicitly or explicitly, the encoder and decoder each maintain
a binary tree, called a code-tree, in which left edges are labelled with
0s, right edges are labelled with 1s and leaves are labelled with the
characters in the alphabet; the binary string on the path from the root to
a leaf is the codeword for that leaf's label.  When the encoder reads a
character $a$, it writes the codeword for $a$ and updates its code-tree;
when the decoder reads the codeword for $a$, it writes $a$ and updates its
code-tree.  Notice that, as long as the encoder and decoder initialize and
update their code-trees according to the same rule, then the encoder's
code-tree when writing a codeword and the decoder's code-tree when reading
that codeword are the same --- so the decoder always writes the same
characters the encoder has read.

\begin{thm}
\label{conversion}
Suppose a number of clients want to send a sequence $S$ of messages, drawn
from a set of size $n$, to a server.  Furthermore, suppose there is a
dynamic instantaneous compression algorithm that, when applied to $S$,
produces a $B$-bit encoding.  Then for any \(k \geq 1\), there is a
dynamic asymmetric communication protocol with which the server sends at
most \(O (n^{1 / k} \log n)\) bits to each client and the clients send, in
total, at most \(k B + 2 |S|\) bits.
\end{thm}

\begin{pf}
The server builds a code-tree $T$ in the same way the encoder would.  For
each message $s$ in $S$, the server truncates $T$ at depth \(\lceil (\log
n) / k \rceil\) to obtain a tree $T'$ (some or all of whose leaves may not
be labelled), then sends $T'$ to the client that has $s$.  Since $T'$ has 
at most \(2^{\lceil (\log n) / k \rceil} \leq 2 n^{1 / k}\) leaves, the 
server can send it to the client in \(O (n^{1 / k} \log n)\) bits.

The client examines $T'$ and, if it finds $s$ labelling a leaf, responds
with 1 followed by the codeword for $s$ according to $T'$ (and $T$); if
not, it responds with 0 followed by the \(\lceil \log n \rceil\)-bit
representation of $s$'s index in the set of possible messages.  Notice
that, in both cases, the server receives enough information to recover
$s$; once it has done this, the server updates $T$ in the same way the
encoder would.  Thus, $T$ is always the same as if maintained by the
encoder.

Let $b$ be the length of the codeword for $s$ in $T$.  If \(b \leq \lceil
(\log n) / k \rceil\), then the client sends \(b + 1\) bits; otherwise,
the client sends \(\lceil \log n \rceil + 1 < k b + 2\) bits.  Since the
compression algorithm encodes $S$ in $B$ bits, all the clients together 
send, in total, at most \(k B + 2 |S|\) bits.
\qed
\end{pf}

\section{Incompressibility and asymmetric communication}
\label{incompressibility}

An advantage of our conversion in Theorem~\ref{conversion}, from dynamic
instantaneous compression to dynamic asymmetric communication, is that the
resulting protocols use only one round of communication for each message;  
i.e., the server sends a truncated code-tree and the client responds with
either a codeword or an index.  Reducing the number of rounds needed is
useful because, as Adler, Demaine, Harvey and
P\v{a}tra\c{s}cu~\cite{ADHP06} wrote:
\begin{quotation}
Any time savings obtained from reducing the number of bits sent by the
client could easily be lost by the extra latency cost induced by multiple
rounds in the protocol, particularly in long-distance networks, such as
satellites, where communication has very high latency.
\end{quotation}
However, in the same paper, they proved a lower bound for protocols that
use few rounds.  They considered the simpler asymmetric communication
problem --- in which the server knows the distribution (with entropy $H$)
over the $n$ possible messages and there is only one client.  Lower bounds
for this simpler problem also hold for dynamic asymmetric communication.  
They proved protocols that use \(o \left( \frac{\log \log n}{\log \log
\log n} \right)\) rounds with high probability and with which the client
is expected to send at most \(O (H + 1)\) bits cannot have a $2^{(\log
n)^{1 - \epsilon}}$ upper bound on the number of bits the server sends,
for any \(\epsilon > 0\).\footnote{This does not contradict the second row
of Table~\ref{results}; Adler and Maggs' second protocol uses \(O (1)\)
rounds in the expected case but not with high probability.}

For the special case of single-round protocols, a stronger, earlier lower
bound was proved by Adler and Maggs~\cite{AM01}: Protocols with which the
client is expected to send at most \(k H\) bits cannot have a
\(\frac{9}{40} n^{1 / (20 k)} \log n\) upper bound on the number of bits
the server sends.  We now prove a nearly tight lower bound for this
special case, assuming transmissions are self-delimiting; we show
afterward that assumption is not necessary.

\begin{thm}
\label{lower_bound}
There does not exist a single-round asymmetric communication protocol with
which, given a probability distribution $P$ with entropy $H$ over a set of
$n$ possible messages, the server sends \(O (n^{1 / k - \epsilon})\) bits
in the worst case and the client sends at most \(k H + o (\log n)\) bits 
in the expected case, for any \(k \geq 1\) and \(\epsilon > 0\).
\end{thm}

\begin{pf}
For the sake of a contradiction, assume there does exist such a protocol.  
Let \(S = s_1, \ldots, s_m\) be a sequence of \(m = n^{1 / k - \epsilon /
2}\) messages chosen uniformly at random and let $P$ be the normalized
distribution of messages in $S$.  By assumption, the server sends sends 
\(O (n^{1 / k - \epsilon}) \subset o (m)\) bits and, by the definition of 
entropy,
\[H
\leq \log m
= \left( \frac{1}{k} - \frac{\epsilon}{2} \right) \log n\ .\]

Let $a'$ denote the client's response when it has message $a$.  Since $P$
is the normalized distribution of the messages in $S$, the length of
\(s_1' \cdots s_m'\) is $m$ times the expected length of the client's 
transmission --- at most
\[m \left( \rule{0ex}{2ex} k H + o (\log n) \right)
= \left( 1 - \frac{\epsilon k}{2} \right) m \log n + o (m \log n)\]
bits.

Since transmissions are self-delimiting, we can store $S$ as the server's
transmission followed by \(s_1' \cdots s_m'\); thus, we can store $S$ in
\(\left( 1 - \frac{\epsilon k}{2} \right) m \log n + o (m \log n)\) bits.  
However, by a simple counting argument, storing $S$ takes \(m \log n\)
bits in the average case.
\qed
\end{pf}

Even if the channel between the server and the client indicates the end of
transmissions --- so the client's response need not be self-delimiting ---
Theorem~\ref{lower_bound} still applies.  To see why, assume there is a
single-round protocol for such channels, with which the server sends \(O
(n^{1 / k - \epsilon})\) bits and the client is expected to send at most
\(k H + o (\log n)\) bits.  We can make all transmissions self-delimiting
by prefacing each transmission by its length encoded in Elias' gamma
code~\cite{Eli75}; in this code, the codeword for the positive integer $i$
has length \(2 \lfloor \log i \rfloor + 1\).  Notice the resulting
protocol still takes a single round; since \(O (\log n^{1 / k} + \log n^{1
/ k - \epsilon}) = O (n^{1 / k - \epsilon})\), the server's
self-delimiting transmission is still \(O (n^{1 / k - \epsilon})\) bits;
by Jensen's Inequality and because \(H \leq \log n\), we have \(k H + 2
\log (k H) + 1 + o (\log n) = k H + o (\log n)\) and the expected length
of the client's self-delimiting transmission is still at most \(k H + o
(\log n)\) bits.  However, Theorem~\ref{lower_bound} forbids the existence
of such a protocol.

\begin{ack}
Many thanks to Faith Ellen, Charles Rackoff and Giovanni Manzini, who 
supervised this research.
\end{ack}

\bibliographystyle{plain}
\bibliography{asymmetric}

\begin{thebibliography}{10}

\bibitem{Adl05}
M.~Adler.
\newblock Collecting correlated information from a sensor network.
\newblock In {\em Proceedings of the 16th Symposium on Discrete Algorithms},
  pages 479--488, 2005.

\bibitem{ADHP06}
M.~Adler, E.D. Demaine, N.J.A. Harvey, and M.~P\v{a}tra\c{s}cu.
\newblock Lower bounds for asymmetric communication complexity and distributed
  source coding.
\newblock In {\em Proceedings of the 17th Symposium on Discrete Algorithms},
  pages 251--260, 2006.

\bibitem{AM01}
M.~Adler and B.M. Maggs.
\newblock Protocols for asymmetric communication channels.
\newblock {\em Journal of Computer and System Sciences}, 64(4):573--596, 2001.

\bibitem{BSTW86}
J.L. Bentley, D.D. Sleator, R.E. Tarjan, and V.K. Wei.
\newblock A locally adaptive data compression scheme.
\newblock {\em Communications of the {ACM}}, 29(4):320--330, 1986.

\bibitem{BKLM03}
P.~Bose, D.~Krizanc, S.~Langerman, and P.~Morin.
\newblock Asymmetric communication protocols via hotlink assignments.
\newblock {\em Theory of Computing Systems}, 36(6):655--661, 2003.

\bibitem{Eli75}
P.~Elias.
\newblock Universal codeword sets and representations of the integers.
\newblock {\em {IEEE} Transactions on Information Theory}, 21:194--203, 1975.

\bibitem{FBHBK02}
N.~Feamster, M.~Balazinska, G.~Harfst, H.~Balakrishnan, and D.~Karger.
\newblock Infranet: Circumventing web censorship and surveillance.
\newblock In {\em Proceedings of the 11th USENIX Security Symposium}, pages
  247--262, 2002.

\bibitem{FBWBK03}
N.~Feamster, M.~Balazinska, W.~Wang, H.~Balakrishnan, and D.~Karger.
\newblock Thwarting web censorship with untrusted messenger discovery.
\newblock In {\em Proceedings of the 3rd International Workshop on Privacy
  Enhancing Technologies}, pages 125--140, 2003.

\bibitem{GGS01}
S.~Ghazizadeh, M.~Ghodsi, and A.~Saberi.
\newblock A new protocol for asymmetric communication channels: Reaching the
  lower bounds.
\newblock {\em Scientia Iranica}, 8(4):297--302, 2001.

\bibitem{Knu85}
D.E. Knuth.
\newblock Dynamic {Huffman} coding.
\newblock {\em Journal of Algorithms}, 6(2):163--180, 1985.

\bibitem{LH02}
E.S. Laber and L.G. Holanda.
\newblock Improved bounds for asymmetric communication protocols.
\newblock {\em Information Processing Letters}, 83(4):205--209, 2002.

\bibitem{Wan03}
W.~Wang.
\newblock Implementation and security analysis of the {Infranet}
  anti-censorship system.
\newblock Master's thesis, Massachusetts Institute of Technology, 2003.

\bibitem{WAF01}
J.~Watkinson, M.~Adler, and F.~Fich.
\newblock New protocols for asymmetric communication channels.
\newblock In {\em Proceedings of the 8th International Colloquium on Structural
  Information and Communication Complexity}, pages 337--350, 2001.

\end{thebibliography}

\end{document}